\documentclass[showpacs,showkeys,preprintnumbers,amsmath,amssymb]{revtex4}

\usepackage{graphicx}

\begin{document}

\title{Electronic Structure of Spin-Chain Compounds: Common Features}


\author{U.\ Schwingenschl\"ogl and C.\ Schuster}


\affiliation{Institut f\"ur Physik, Universit\"at Augsburg, 86135 Augsburg,
Germany}


\begin{abstract}
The incommensurate composite systems M$_{14}$Cu$_{24}$O$_{41}$ (M=Ca,Sr,La)
are based on two fundamental structural units: CuO$_2$ chains and
Cu$_2$O$_3$ ladders. We present electronic structure calculations
within density functional theory in order to address the interrelations
between chains and ladders. The calculations account for the
details of the crystal structure by means of a unit cell comprising 10
chain and 7 ladder units. It turns out that chains and ladders can be treated
independently, which allows us to introduce a model system based on a
reduced unit cell. For the CuO$_2$ chains, we find two characteristic
bands at the Fermi energy. Tight binding fits yield
nearest and next-nearest neighbour interactions of the same order of
magnitude.
\end{abstract}

\maketitle

The isostructural spin-chain compounds M$_{14}$Cu$_{24}$O$_{41}$
(M=Ca,Sr,La) have been subject of intensive research in recent years,
mainly due to their rich phase diagram and close relations to the
high-T${\rm_c}$ cuprates. Their incommensurate crystal structures consist
of planes of quasi one-dimensional CuO$_2$ chains stacked alternately with
planes of two-leg Cu$_2$O$_3$ ladders. The orientation of these chains and
ladders defines the crystallographical $c$-axis, where the lattice
constants of the two subsystems satisfy in a good approximation
$10c_{\rm chain}\approx7c_{\rm ladder}$ \cite{ukei94,gotoh03}. The
copper ions are intrinsically hole doped with nominal Cu valence +2.25
for both the Ca and the Sr compound. However, the ladders contain less
holes than the chains. The resulting Cu valence of +2.75 for the
chain sites corresponds to a quarter filled one-dimensional band
\cite{nuecker00}. Optical conductivity as well as x-ray absorption
experiments suggest that substitution of Ca for Sr induces a transfer of
holes from the chains back to the ladders \cite{osafune97}. On La
substitution the intrinsic doping decreases, reaching the undoped state
with nominal Cu valence +2.00 for La$_6$Ca$_8$Cu$_{24}$O$_{41}$.

In the case of Sr$_{14}$Cu$_{24}$O$_{41}$ the CuO$_2$ chains are
non-magnetic with a spin gap of about 130\,K \cite{matsuda96} and the
Cu$_2$O$_3$ ladders show a charge density wave in a wide temperature range
\cite{gorshunov02, gozar03}. In contrast, for Ca rich samples
antiferromagnetic ordering is found in the chains \cite{nagata99,isobe00}
and the ladders become superconducting under pressure \cite{uehara96}.
This strong doping dependence of both the charge and the magnetic order
\cite{carter96,ammerahl00,braden04,kataev01} is accompanied by only minor
modifications of the crystal structure \cite{ohta97,matsuda97,klingeler03}.
From the theoretical point of view, the magnetic properties of undoped
two-leg S=1/2 ladders have been analyzed be means of a Heisenberg model
including cyclic ring exchange \cite{nunner02}. Possible effects of hole
doping are investigated in \cite{nishimoto02}. A large spin gap makes the
Cu$_2$O$_3$ ladder planes magnetically inert. On the other hand, the magnetic
phase diagram of the CuO$_2$ chains is quite complicated. Ferromagnetism,
as expected for Cu-O-Cu bond angles of approximately 90$^\circ$ in the
chains, is realized in La rich systems. The origin of the observed
intrachain antiferromagnetic order for a quarter filled Cu band, however,
led to controversial discussions in the literature
\cite{klingeler03,gelle05,klingeler06}. Understanding of the doping
dependence of both the charge order and the spin gap is likewise far from
complete.

Electronic structure calculations applying density functional theory
and the local density approximation have been performed by M\"uller
{\it et al.} \cite{mueller98} for the isostructural Cu$_2$O$_3$ ladders in
SrCu$_2$O$_3$ and by Arai {\it et al.} \cite{arai97} for a simplified
Sr$_{14}$Cu$_{24}$O$_{41}$ unit cell. Probably because of a high demand on
CPU time, state-of-the-art band structure calculations accounting for
the details of the crystal structures are absent so far. In the present
letter we aim at filling this gap in order to address the interplay between the
chain and ladder subsystems and to identify
electronic features common to all compounds due to the specific crystal
structure. Moreover, we introduce a model system capturing the
relevant properties of the CuO$_2$ chains for various compounds and
levels of hole doping. By this model system, we identify the
electronic states at the Fermi level and study intrachain nearest and
next-nearest neighbour interactions in the framework of a tight binding
model. As we account for the decisive structural ingredients, we obtain,
as far as we know for the first time, realistic insights into the electronic
structure of the CuO$_2$ chains in spin-chain compounds.

Our first principles band structure calculations for the full
Sr$_{14}$Cu$_{24}$O$_{41}$ unit cell are based on the scalar-relativistic
augmented spherical wave (ASW)
method \cite{eyert00}, which has proven to be suitable for dealing
with unit cells comprising a large number of atomic sites
\cite{us03,us04,us07}. As structural input we apply the crystallographical
data due to a recent structural refinement by Gotoh {\it et al.}
\cite{gotoh03}. A previous investigation of the crystal structure by Ukei
{\it et al.} \cite{ukei94} seems to suffer from low quality samples. Since the
Sr$_{14}$Cu$_{24}$O$_{41}$ unit cell contains two formula units, we have
to deal with 28 strontium, 48 copper, and 82 oxygen atomic spheres.
Furthermore, 146 additional augmentation spheres are necessary in order to
represent the correct shape of the crystal potential in the large voids
of the open crystal structure. The basis sets taken into account in the
secular matrix therefore comprise Sr $5s$, $5p$, $4d$, Cu $3d$, $4s$, $4p$, 
and O $2s$, $2p$ orbitals, as well as states of the additional augmentation
spheres. Brillouin zone integrations are performed using an increasing
number of finally 72 {\bf k}-points in the irreducible wedge. The
Vosko-Wilk-Nusair parametrization is used for the exchange-correlation
functional.

The model system for the CuO$_2$ chain discussed subsequently comprises
6 atomic sites. It is studied via band
structure calculations using the WIEN2k program package, which is a popular
full-potential linearized augmented plane wave (LAPW) code \cite{wien2k}.
To obtain reliable results, we apply basis sets of 18135 plane
waves and {\bf k}-meshes with up to 108 points in the irreducible
wedge of the Brillouin zone. Whereas the Cu $3p$ and O $1s$ orbitals are
treated as semi-core states, the valence states consist of Cu $3d$, $4s$,
$4p$ and O $2s$, $2p$ orbitals. The data are based on the
Perdew-Burke-Ernzernhof scheme for the exchange-correlation functional.

\begin{figure}[t!]
\includegraphics[width=70mm]{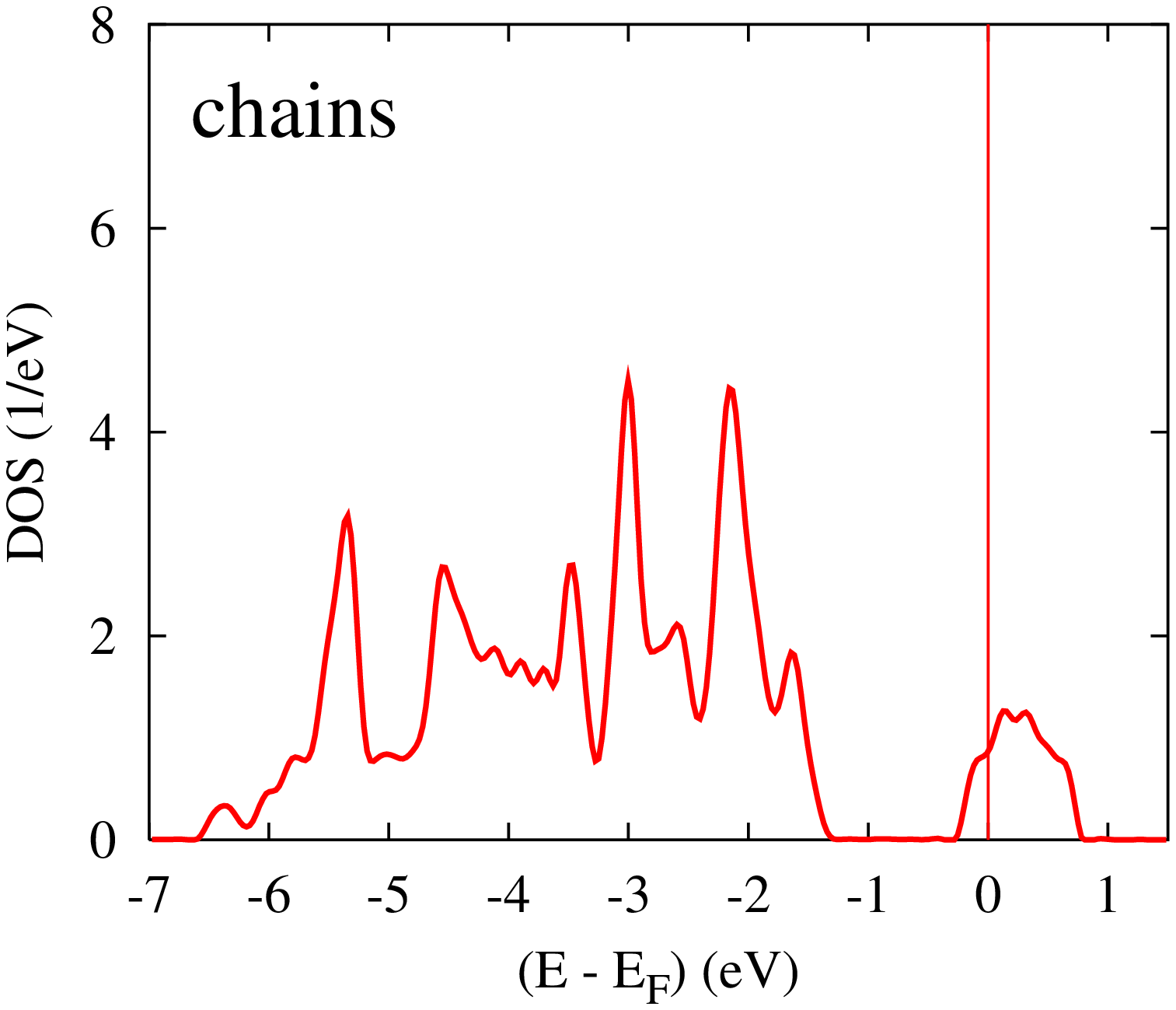}\includegraphics[width=70mm]{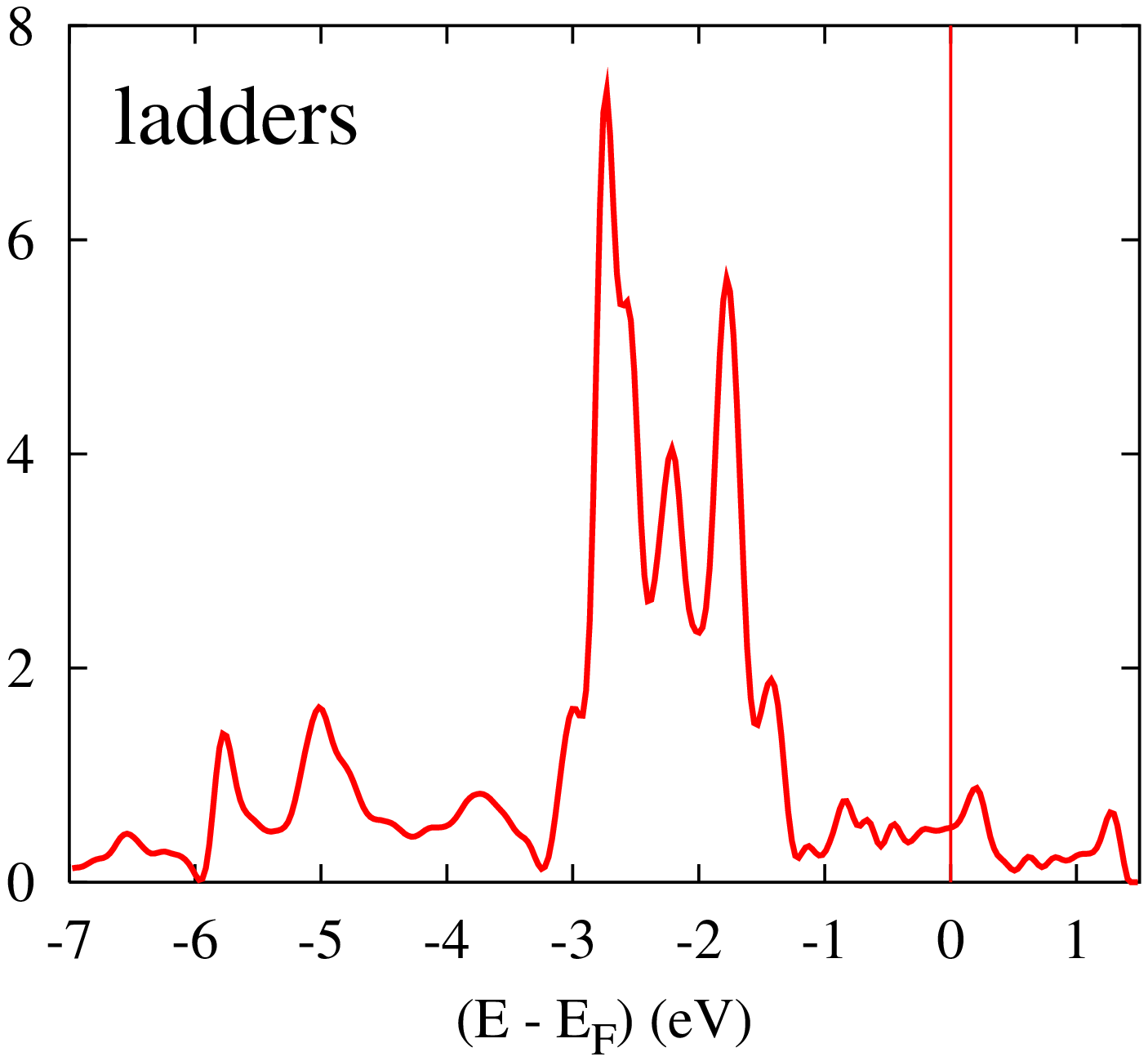}\\
\includegraphics[width=70mm]{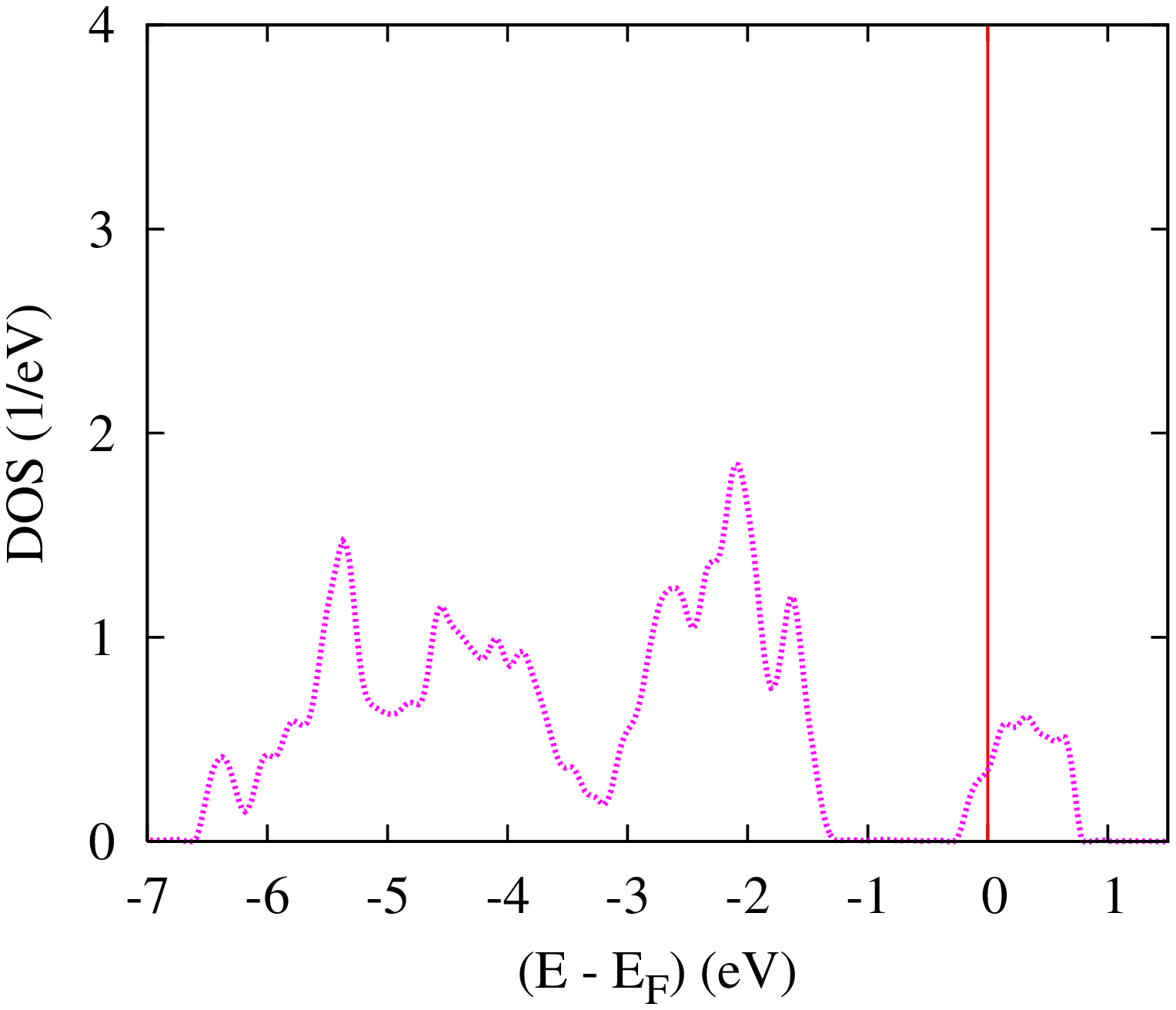}\includegraphics[width=70mm]{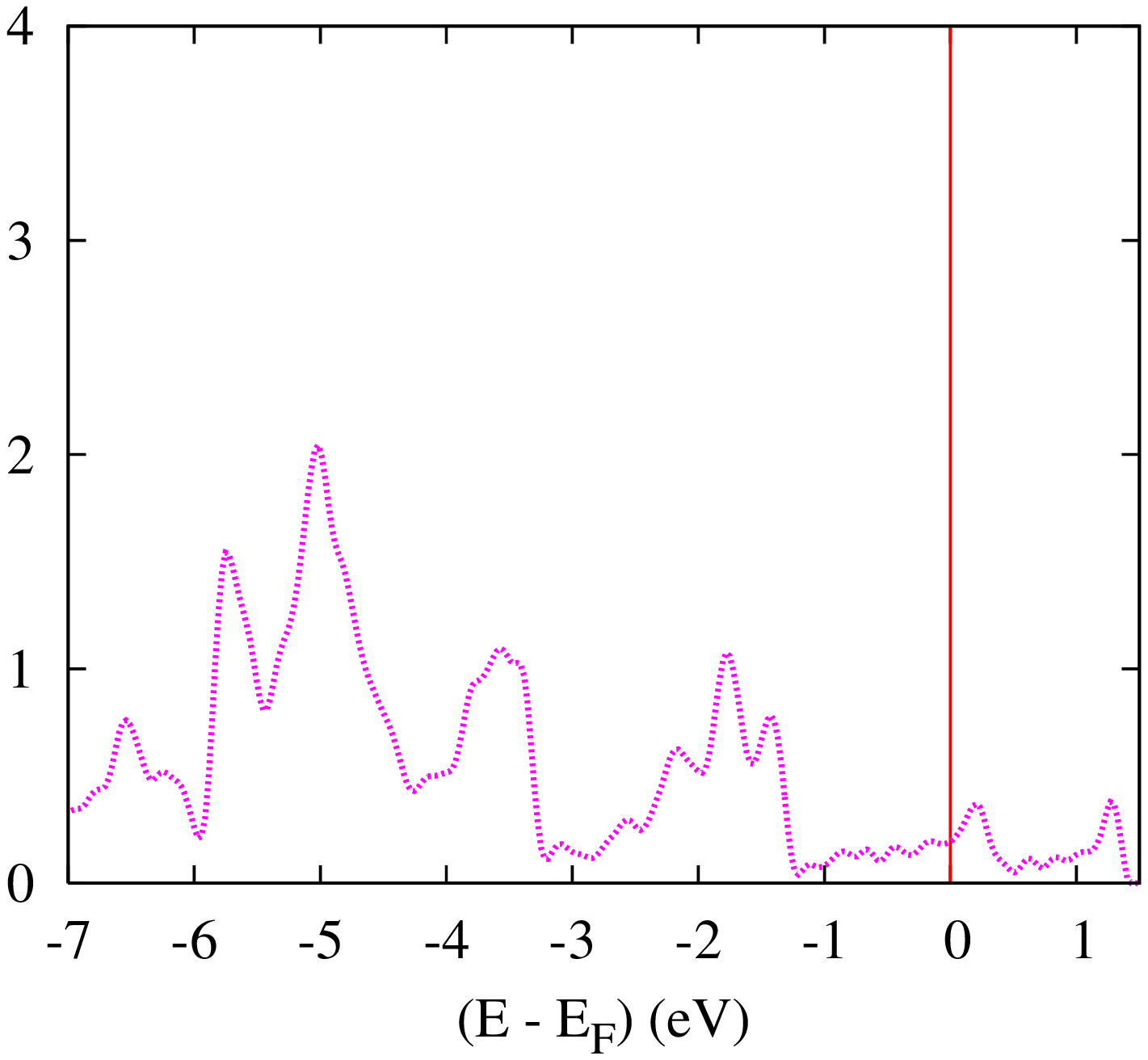}
\caption{Partial Cu $3d$ (first row) and O $2p$ (second row) densities
of states (per atom) for the chain (left) and ladder (right) subsystems.
The results are based on the full Sr$_{14}$Cu$_{24}$O$_{41}$ unit cell.}
\label{fig1}
\end{figure}

Figure \ref{fig1} displays partial Cu $3d$ densities of states (DOS)
resulting from our calculations for the full Sr$_{14}$Cu$_{24}$O$_{41}$
unit cell. The Cu $3d$ DOS is seperated into contributions due to the
CuO$_2$ chains (left) and Cu$_2$O$_3$ ladders (right), and normalized with
respect to the number of contributing sites. We take into account the
structural details of the compound by means of a unit cell comprising 10
chain and 7 ladder units along the crystallographical $c$-axis. This
unit cell gives rise to an excellent approximation to the incommensurate
crystal structure \cite{gotoh03}. Note that different lengths of the chain
and ladder units in principle lead to slight periodical tiltings, which do
not affect our conclusions and therefore are not considered here.

\begin{figure}[t!]
\includegraphics[width=70mm]{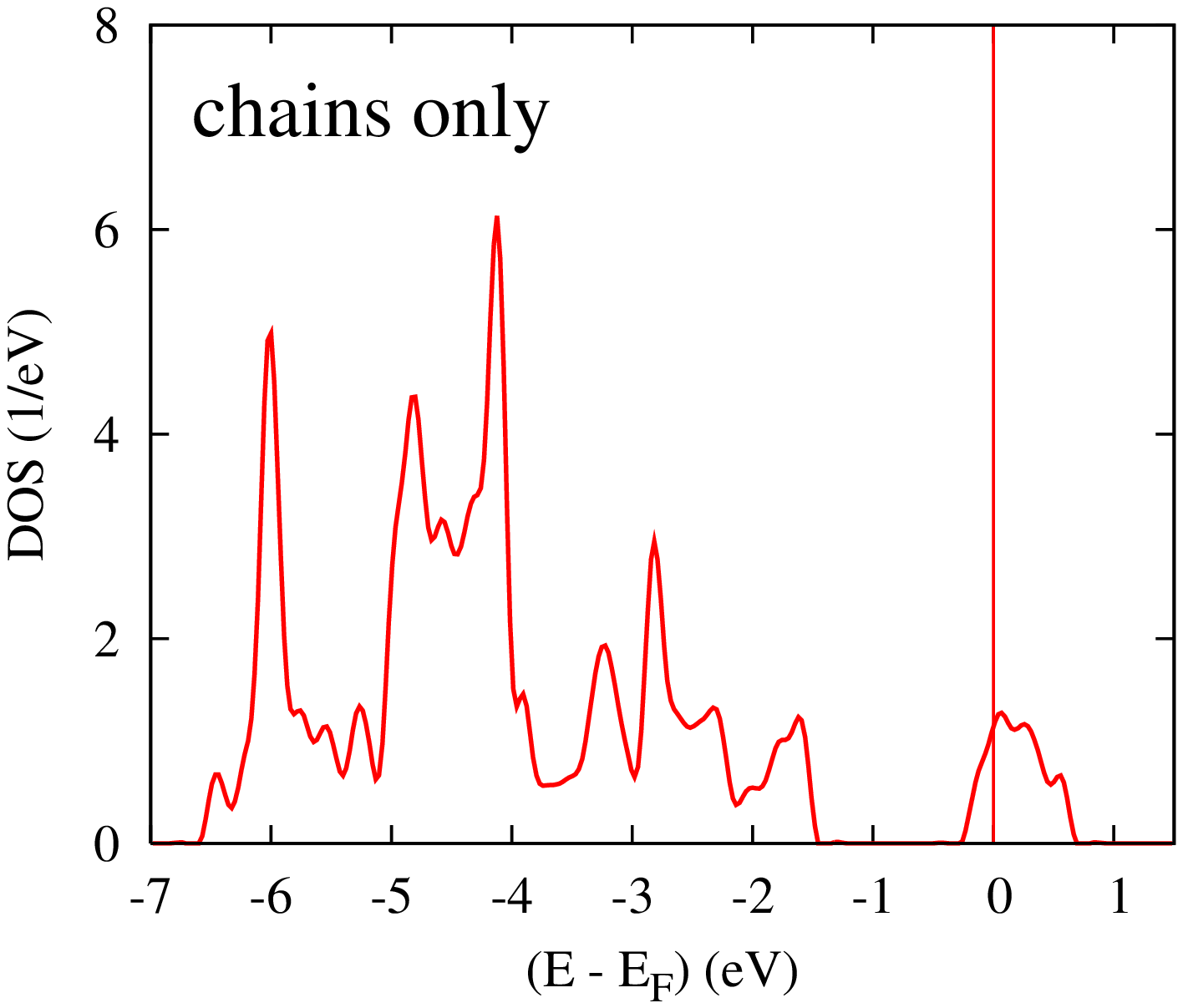}
\includegraphics[width=70mm]{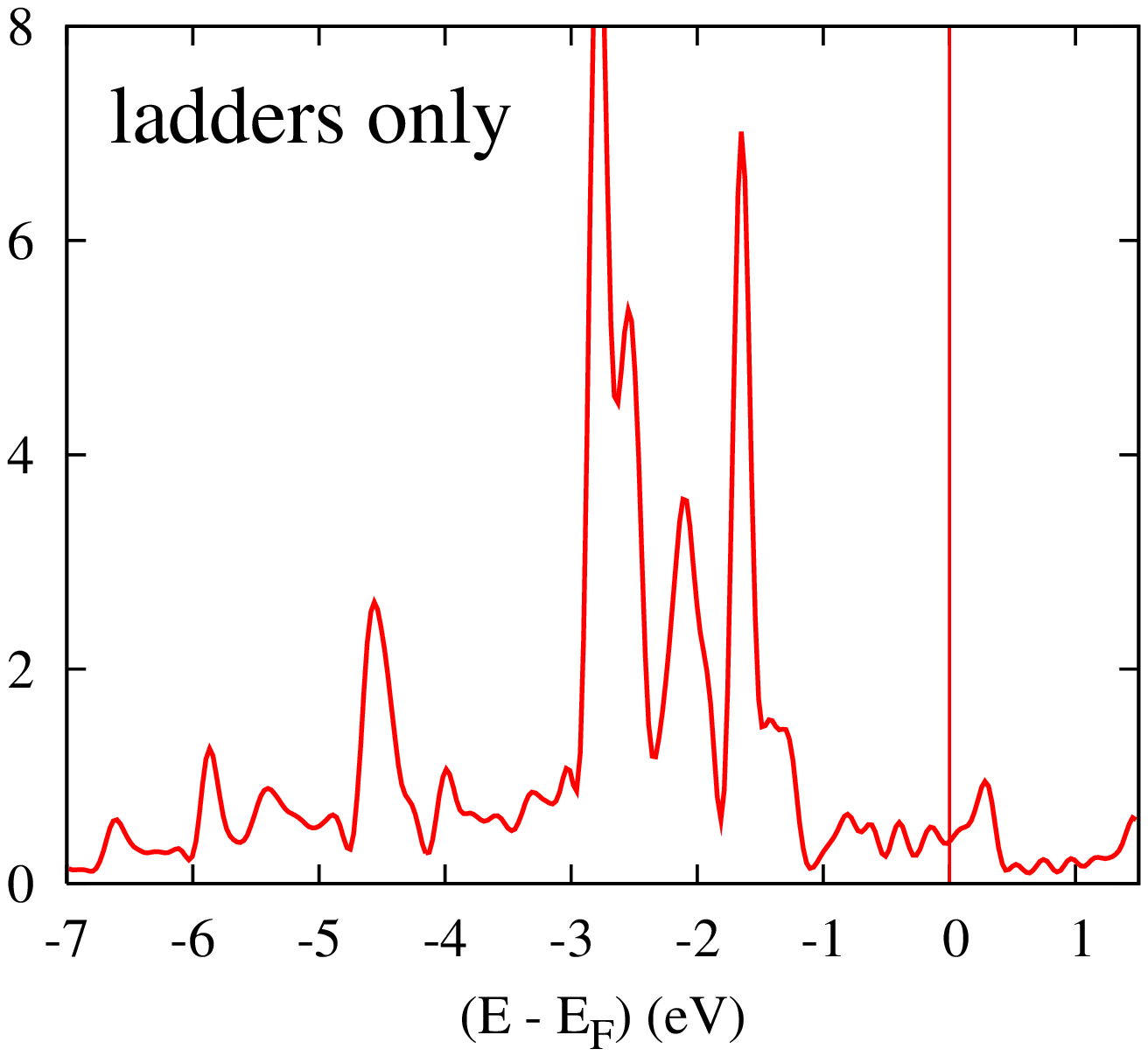}
\caption{Partial Cu $3d$ densities of states for chain (left) and
ladder (right) copper sites. The results are based on
Sr$_{14}$Cu$_{24}$O$_{41}$ unit cells where either the chain or the
ladder substructure was removed. As concerns the vicinity of the Fermi
level, the curves resemble almost perfectly the findings for the chain
and ladder copper sites in the full Sr$_{14}$Cu$_{24}$O$_{41}$ unit cell,
respectively, compare fig.\ \ref{fig1}.}
\label{fig2}
\end{figure}

Both for the chains and ladders there is a finite DOS at the Fermi energy
in fig.\ \ref{fig1}, contradicting the experimentally observed non-metallic
state of Sr$_{14}$Cu$_{24}$O$_{41}$. In order to reproduce the insulating
behaviour, it would be necessary to include electronic correlations beyond
the local density approximation. However, for our purpose of studying the
interplay between the chains and ladders such local interactions are not of
interest. Electronic states in the energy range of fig.\ \ref{fig1} are
composed of Cu $3d$ and O $2p$ orbitals (almost exclusively), which
indicates strong Cu-O hybridization both in the chains and in the ladders.
As expected, this resembles the findings for the simplified unit cell of
Arai {\it et al.} \cite{arai97}. The same is true for the gross shapes of
the chain and ladder DOS curves. While the ladder DOS is spread over the
whole energy range from $-7$\,eV to $1.5$\,eV, the chain DOS shows a
distinct band with a width of 1\,eV near the Fermi level, separated by a
gap of likewise 1\,eV from the other valence bands. This band
is close to quarter filling, as to be expected, and hence subject to a large variety of
possible ordering processes of the charge and spin degrees of freedom
\cite{schuster07}.

\begin{figure}[t!]
\includegraphics[width=70mm]{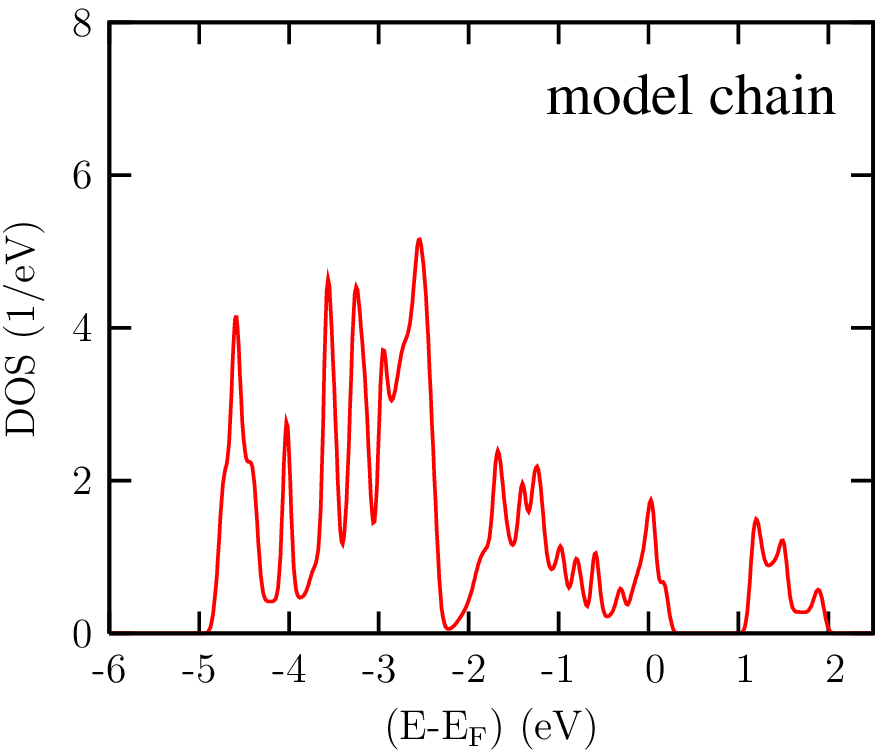}
\includegraphics[width=70mm]{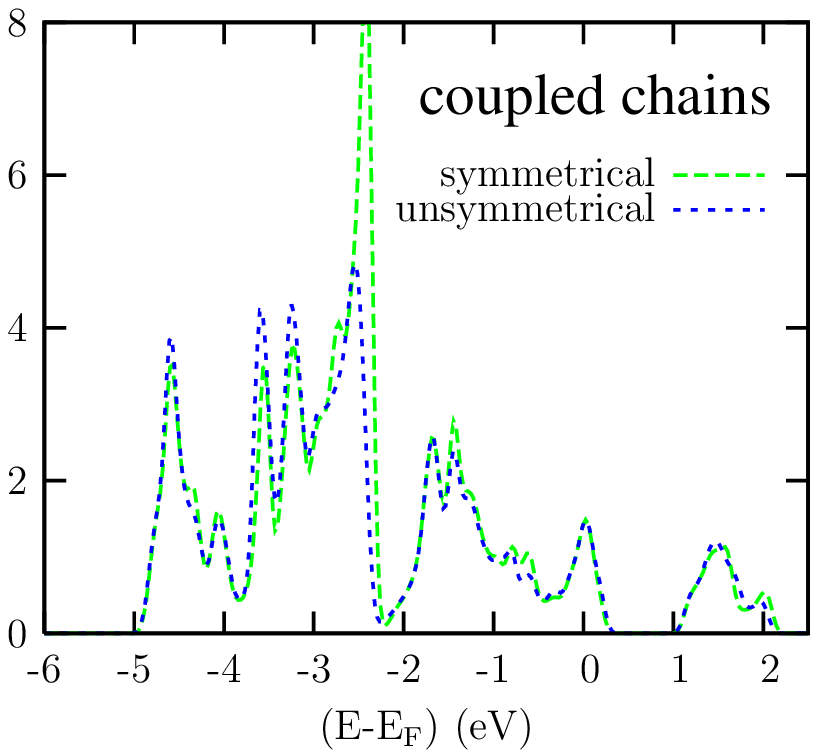}
\caption{Partial Cu $3d$ densities of states for a model CuO$_2$ chain
(left) and two coupled CuO$_2$ chains (right), arranged symmetrically
and unsymmetrically. The states are subject to a
rigid band shift of about 1.5\,eV as compared to the real system.}
\label{fig3}
\end{figure}

\begin{figure}[t!]
\includegraphics[width=70mm]{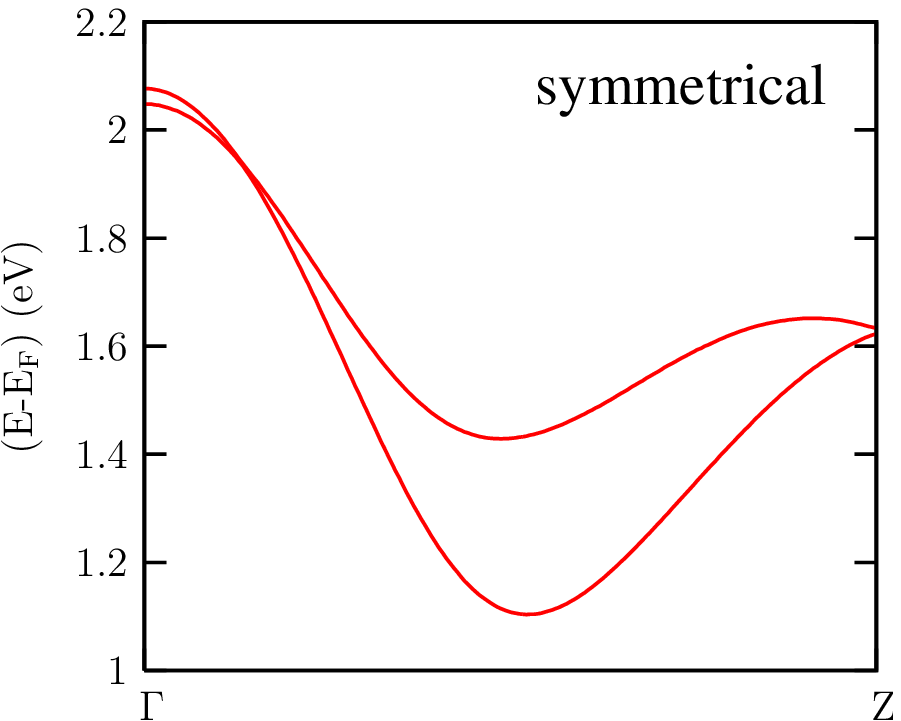}\includegraphics[width=70mm]{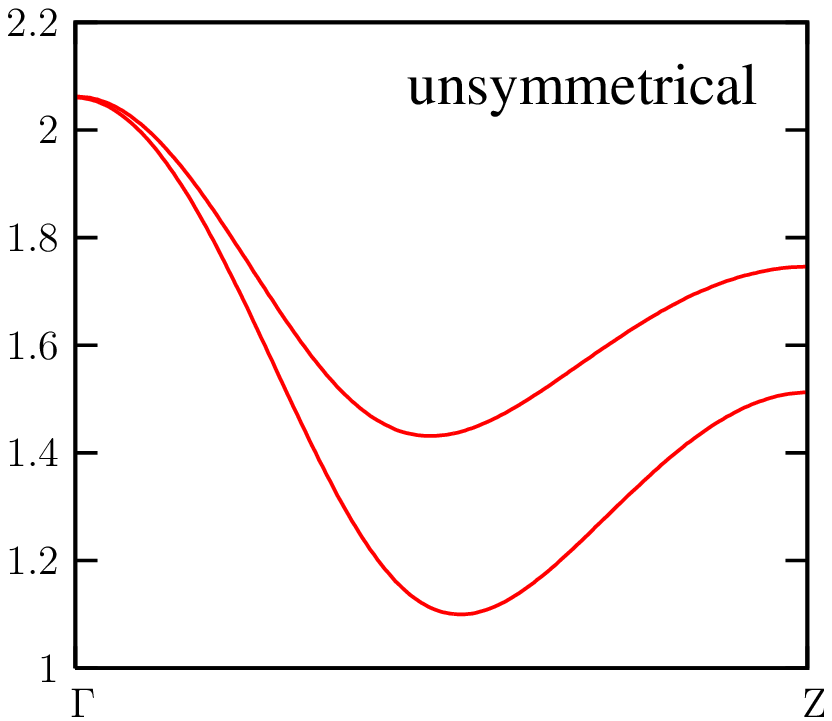}
\caption{Electronic band structure (Cu $3d$ bands) for coupled CuO$_2$
chains, arranged symmetrically (left) and unsymmetrically (right). The
results of tight-binding fits are summarized in tabel \ref{tab1}.}
\label{fig4}
\end{figure}

Turning to the question about the interrelations between the chain and
ladder subsystem, the Sr$_{14}$Cu$_{24}$O$_{41}$ unit cell is modified
as follows: (a) We remove either the chain or the ladder substructure
and (b) replace Sr by K. Since the M ions in
M$_{14}$Cu$_{24}$O$_{41}$ act almost exclusively as charge donors, the
latter allows us to maintain the correct doping when removing one of the
substructures. For the full unit cell, formally one electron is
transferred from each Sr site to both the chain and the ladder subsystem.
The same amount of charge transfer is achieved for the half-filled unit
cell via the potassium, because hybridization effects between Sr/K and
Cu/O orbitals are negligible. Figure \ref{fig2} displays DOS curves for
the chains only configuration on the left and for the ladders only
configuration on the right side. In both cases the findings resemble the
corresponding curves of fig.\ \ref{fig2} as concerns the total width of
the valence bands as well as the gross shape of the DOS. In the vicinity
of the Fermi energy even the details of the DOS are identical. In
particular, for the CuO$_2$ chains, the band width and the filling of the
distinct band around the Fermi level hardly alter.

To conclude, since the above argumentation does not depend on the
specific choice of the donor ion M, we find that spin-chain compounds
M$_{14}$Cu$_{24}$O$_{41}$ are well described in terms of three largely
independent subsystems: The CuO$_2$ chains, the Cu$_2$O$_3$ ladders,
and the electron donor system of M ions, which separates chains and ladders.
Hybridization between atomic orbitals belonging to different
subsystems can be neglected because the valence states of the M ions are
almost fully depopulated. As a consequence, we attribute charge transfer
between chains and ladders when replacing Sr by isoelectronic Ca
\cite{osafune97,nuecker00} to simultaneous modifications of the crystal
structure.

\begin{table}
\begin{tabular}{l|c|c|c|c}
&band 1, symm.& band 2, symm.&band 1, unsymm.&band 2, unsymm.\\\hline
$t_1$ (eV)&-0.115&-0.105&-0.135&-0.080\\
$t_2$ (eV)&-0.183&-0.098&-0.165&-0.115\\
$\epsilon_0$ (eV)&1.485&1.645&1.460&1.675
\end{tabular}
\vspace*{0.5cm}
\caption{\rm Tight-binding parameters for the Cu $3d$ bands of
fig.\ \ref{fig4}.}
\label{tab1}
\end{table}

Since we have established that the structural subsystems of spin-chain
compounds can be treated independently, we address the electronic
features of the CuO$_2$ chains in the following via a model system
based on a reduced unit cell. For convenience, we choose the Euklidean
$z$-axis as the chain axis and use the interatomic distances and bond
angles of Sr$_{14}$Cu$_{24}$O$_{41}$. Each Cu site thus is connected
to two symmetrical O sites in both chain directions, where the bond
lengths are 1.89\,\AA\ in one and 1.87\,\AA\ in the other direction. The
related O-Cu-O bond angles are 85.5$^\circ$ and 86.7$^\circ$,
respectively. Neighbouring Cu atoms are separated by 2.75\,\AA. In order
to prepare for the symmetry analysis of the Cu bands, we align the CuO$_2$
chains in the Euklidean $xz$-plane. As a consequence, Cu $3d_{xz}$
orbitals mediate the main part of the Cu-O overlap. We study both
a single CuO$_2$ chain, for which the unit cell comprises just one
CuO$_2$ unit, and two coupled chains. The latter allows us to address
the influence of relative shifts between adjacent chains with respect to
the chain axis, which distinguish specific spin-chain compounds.
For Ca$_{13.6}$Sr$_{0.4}$Cu$_{24}$O$_{41}$ \cite{isobe00,ohta97} the
chains are shifted by half their intrachain Cu-Cu distance, for example,
while the shift is only 30\% of the Cu-Cu distance for
Sr$_{14}$Cu$_{24}$O$_{41}$. We compare these two cases, which we call
the symmetrically and unsymmetrically arranged chains,
respectively.

Figure \ref{fig3} summarizes the results of our LAPW model calculations,
showing partial Cu $3d$ densities of states for a single CuO$_2$ chain
on the left side and two coupled CuO$_2$ chains on the right side.
All three DOS curves resemble the essential features of the ASW data
for the real system, see figs.\ \ref{fig1} and \ref{fig2}. We obtain a
widespread structure with a width of about 5.2\,eV at lower energies and
a structure extending over a range of 1\,eV at higher energies. However,
comparison with the ASW findings results in a rigid band shift of slightly
less than 1.5\,eV to higher energies. This shift traces back to the
fact that the M ions are not taken into account in the model system,
see the previous discussion. In the following
we hence are interested in the electronic states in the energy range from
about 1\,eV to 2\,eV. As expected, these states are mainly due to the
antibonding combination of Cu $3d_{xz}$ and both O $2p_x$ and $2p_z$
orbitals. Contributions of other states amount to less than 0.1\% of the
total DOS at the Fermi energy. In order to obtain further insight into
the electronic features of the CuO$_2$ chains, we turn
to the band structure underlying the DOS curves of fig.\ \ref{fig3}.

Because the CuO$_2$ chains run parallel to the $z$-axis of our unit cell,
fig.\ \ref{fig4} depicts the Cu $d_{xz}$ bands in the corresponding
direction of the orthorhombic Brillouin zone, i.e.\ between the
high symmetry points $\Gamma$ and $Z$. As the length of a CuO$_2$ unit
amounts to $c_{\rm chain}$=2.75\,\AA, we can write
${\bf k}_\Gamma=(0,0,0)$ and ${\bf k}_Z=(0,0,2\pi/c_{\rm chain})$.
Reasonable fitting of the bands is possible by assuming a the
tight-binding dispersion of the form
\begin{equation}\label{eq1}
\epsilon(k_z)=
\epsilon_0-2t_1\cos(k_zc_{\rm chain})-2t_2\cos(2k_zc_{\rm chain}),
\end{equation}
where $k_z$ is the $z$-component of the reciprocal lattice vector and
$\epsilon_0$ is the band center. Moreover, $t_1$ and $t_2$ denote the
nearest and next-nearest neighbour hopping parameters along the chains,
respectively. The outcome of the tight-binding fits for the bands of
fig.\ \ref{fig4} is given in table \ref{tab1}. Beyond our one-dimensional
model (\ref{eq1}), influence of finite interchain hopping can be
identified in the band structure for the symmetrical configuration, but
is not relevant for our further conclusions. In contrast to Arai
{\it et al.} \cite{arai97}, we find two Cu $3d_{xz}$ bands, corresponding
to the two coupled chains. We attribute this difference to inadequate
crystallographical data entering the former band calculation. Importantly,
we obtain for both Cu $3d_{xz}$ bands nearest and next-nearest neighbour
couplings of the same order of magnitude, independent of shifts between
neighbouring CuO$_2$ chains. Despite the 90$^\circ$ Cu-O-Cu intrachain
bond angles, strong nearest neighbour interaction therefore appears to be
typical of the whole class of spin-chain compounds. As the calculated
magnitudes of the nearest and next-nearest neighbour coupling are common
to all systems, we conclude that changes in the magnetic ordering on
replacing Ca by La result almost exclusively from the modified filling
of the Cu $3d_{xz}$ bands.

In summary, we have studied the electronic structure of
Sr$_{14}$Cu$_{24}$O$_{41}$ by means of density functional theory. Taking
into account the details of the crystal structure, we have addressed the
interrelations between the structural subsystems the spin-chain compounds
M$_{14}$Cu$_{24}$O$_{41}$ (M=Ca,Sr,La) are composed of. Hybridization
between the CuO$_2$ chains, the Cu$_2$O$_3$ ladders, and the electron
donor M ions is found to be negligible. Since these subsystems consequently
can be treated individually, we have analyzed a simplified model system
for the CuO$_2$ chains to identify features common to all spin-chain
compounds. It turns out that two characteristic Cu $3d_{xz}$ bands at the
Fermi energy dominate the electronic and magnetic properties of the chains.
The filling of these bands distinguishes different compounds.
Tight-binding fits prove that nearest and next-nearest neighbour
interactions in the CuO$_2$ chains are of the same order of magnitude,
which applies to the whole class of spin-chain systems.

\begin{acknowledgments}
We thank U.\ Eckern for helpful discussions and the
Deutsche Forschungsgemeinschaft for financial support (SFB 484). 
\end{acknowledgments}

\end{document}